# Study on 2015 June 22 Forbush decrease with the muon telescope in Antarctic *

De-Hong Huang (黄德宏)[1,1)]   Hong-Qiao Hu(胡红桥)[1]   Ji-Long Zhang (张吉龙)[2,2)]
Hong Lu (卢红)[2]   Da-Li Zhang (张大力)[2]   Bin-Shen Xue[3]   Jing-Tian Lu[3]

[1] (SOA Key Laboratory for Polar Science, Polar Research Institute of China, Shanghai 200136, China

[2] Institute of High Energy Physics (IHEP), Chinese Academy of Sciences (CAS), Beijing 100049, China

[3]National Satellite Meteorological Center, Beijing 10081, China

Abstract: By the end of 2014, a cosmic ray muon telescope was installed at Zhongshan Station in Antarctic and has been continuously collecting data since then. It is the first surface muon telescope to be built in Antarctic. In June 2015, five CMEs were ejected towards the Earth initiating a big large Forbush decrease (FD) event. We conduct a comprehensive study of the galactic cosmic ray intensity fluctuations during the FD using the data from cosmic ray detectors of multiple stations (Zhongshan, McMurdo, South Polar and Nagoya) and he solar wind measurements from ACE and WIND. A pre-increase before the shock arrival was observed. Distinct differences exist in the timelines of the galactic cosmic ray recorded by the neutron monitors and the muon telescopes. FD onset for Zhongshan muon telescope is delayed (∼2.5h) with respect to SSC onset. This FD had a profile of four-step decrease. The traditional one- or two-step classification of FDs was inadequate to explain this FD.

## 1   Introduction

Decreases in the cosmic ray count rate which last typically for about a week , were first observed by Forbush (1937)[1] and Hess and Demmelmair (1937)[2] using ionization chambers. It was the early 1960s work of Simpson using neutron monitors (|Simpson, 1954) [3] which showed that the origin of these decreases was in the interplanetary medium.

Forbush decrease (FD) is a transient depression in the galactic cosmic ray (GCR) intensity which is typically characterized by a sudden onset, reaching a minimum within about a day, followed by a more gradual recovery phase typically lasting for several days. The magnitude of FD varies from a few percent up to 25 % in the neutron monitor energy range. Although FD are known as suppressions of the ground-bases detector count rates since long, direct measurements of GCR energy spectra during FD were not performed until the launch of PAMELA instrument in 2006 [4].

 FDs are usually caused by transient interplanetary events, which are related to coronal mass ejections (CMEs) from the Sun, and also the corotating interacting regions (CIRs) between the fast and slow solar wind streams from the Sun. The near-Earth manifestations of CMEs from the Sun typically have two major components: the interplanetary counterpart of the CME, commonly called an ICME, and the shock which is driven ahead of it. ICMEs which possess certain well-defined criteria such as

*Supported by NSFC (11575204)

1)E-mail:huangdehong@pric.org.cn

2) E-mail: zhangjl@ihep.ac.cn

reductions in plasma temperature and smooth rotations of the magnetic field are called magnetic clouds, while others are often classified as ejecta. Thus CME-related cosmic ray decrease are of three basic types: those caused by a shock and ejecta, those caused by a shock only and those cause by an ejecta only[5]. The relative contribution of ICMES and shocks in causing FD is a matter of debate [6]. In term of understanding the internal magnetic topology of CMEs in the interplanetary medium, cosmic ray anisotropies should provide valuable information which cannot be obtained by any other type of in situ measurement.

Cosmic-ray observation from satellites has a unique advantage: direct measurement of cosmic rays while avoiding the atmospheric effect; however, the small size of satellite-based detectors does not provide sufficient statistics in the high-energy region for real-time monitoring of the space environment. For instance, the AMS onboard ISS, measures particles in the GV-TV rigidity range, only provide high energy Solar Energetic Particles (SEPs) events at ∼ 1 GV from 2011 to 2016[7]. On the other hand, cosmic ray detectors on the ground can monitor the intensity of cosmic rays from 500 MeV to 100 GeV with sufficient statistics. They have been gathering important data on the space environment continuously for many years. Space and ground observations should be conducted simultaneously in such away so as to supply complementary data for a full understanding of the space environment.

Since 1957, ground observation have been carried out by means of NM-64 neutron monitors: Climax (USA), Oulu (Finland), Hermanus (South Africa), Irkutsk (Russia), Jungfraujoch (Switzerland), Kiell (Germany), Lomnicky (Slovakia), Moscow (Russia), Mt. Norikura (Japan), Rome (Italy), Thule (Greenland), Yangbajing (China). The muon detectors complement neutron monitors by monitoring the cosmic ray modulation at slightly higher energy and provide measurements of numerous arrival directions of muons from a single location. A typical energy for a muon detector directional channel would be ∼50 GeV, as compared to ∼10 GeV for neutron monitors. In recent years, some new ground-based muon monitors have been set up around the world to perform space environment studies using cosmic rays. Global Muon Detector Network (GMDN) was established in March 2006, when a hodoscope type cosmic ray detector (9 $m^2$) in Kuwait was added to the previous network composed of the multidirectional detectors (36 $m^2$) in Nagoya, Japan, a prototype muon detector (4$m^2$, expansion to 36$m^2$ in 2012)in Sao Martinho, Brazil, and a smaller one (9$m^2$ ) in Hobart, Australia [8].

A new muon-neuton telescope was completed in Yangbajing in 2007 [9]. The Yangbajing Cosmic Ray Observatory is locate at latitude and longitude coordinates of 30.11°N and 90.53°E, respectively, at an altitude of 4300 m above sea level. The vertical cutoff rigidity for cosmic rays is 14.1 GV.

By the end of 2014, a cosmic ray muon telescope was installed at Zhongshan Station in Antarctic [10]. It is locate latitude and longitude coordinates of 69.4°S and 76.4°E, respectively, at sea level. The vertical cutoff rigidity of cosmic ray protons at Zhongshan Station is 0.76 GV. The observation data is sent to China by satellite link and is released through network. Every 1 hour, the raw data of the muon telescope are transmitted from Zhongshan station to a server belonging to IHEP (Institute of High Energy Physics) in Beijing by satellite link. The raw data are analyzed and stored in a MySQL library. The server then publishes real-time data on the Internet. Users can either view the data graphically or download data (of resolution from every 60 seconds to daily averages) using a web browser from the address [11].

In June 2015, five CMEs were ejected towards the Earth initiating a big large FD event. We examine the sequence of solar events that occurred during 21-25 June and how they affect the counting rates of the Zhongshan muon telescope. A comparative analysis of this FD event with Zhongshang, McMurdo, South Poslar and Nagoya is performed. We present, for the first time, a surface muon telescope in Antarctic records the FD event.

This paper consists as follows. In section 2, we briefly describe the sitting of the Zhongshan Muon Telescope. The observed results for FD are discussed in Section 3, and summary are drawn in Section 4.

## 2 The Zhongshan Muon Telescope

By the end of 2014, a cosmic ray muon telescope was installed at Zhongshan Station. we briefly describe the setting of the Zhongshan Muon Telescope. Meteorological effects of Zhongshan Station muon telescope observational data are discussed.

### 2.1 Setting

The Zhongshan Muon Telescope (ZSMT) consists of two plastic scintillator detectors. Each scintillator is rectangular 50 cm × 50 cm and 2 cm thick. Those two scintillators are vertically separated by 75 cm and can determine the arrival direction of muons. A photomultiplier tube (PMT) was placed 50 cm above the scintillator in to obtain a high uniformity of light collection. The output pulse from a PMT is amplified and shaped into a digital signa. When a particle passed through the upper and lower layers, an output signal from each detector is sent to the logic modules and a coincidence trigger pulse is generated by a FPGA (Field Programmable Gate Array). The trigger pulses are recorded by a scaler. Every 1 s, the scaler data is send to the PC computer via an RS2esC serial port. All these circuits are realized by one chip, the FPGA (Xilinx SPARTAN XC3S250E). For the convenience of conveying and installing, the lead plate between the two scintillators was not installed.

### 2.2 Meteorological effects of Zhongshan Station muon telescope observational data

Figure 1 shows hourly cosmic-ray counts and atmosphere pressure observed by the ZSMT, from the 1st to 30th of June 2015. It is observation that the pressure change is very strong at Zhongshan station. This change is from 963.85 hPa to 1004.93 hPa and change value is about 41 hPa. During the 1st to 30th of June 2015, the change is from 596.78 hPa to 607.48 and change value is about 10 hPa at Yangbajing station, Tibet. It is necessary to a pressure correct of muon data for ZSMT.

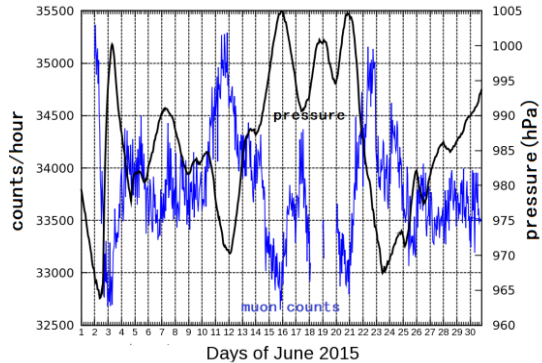

Fig. 1 Meteorological effect of ZSMT data

From observed data from 1st to 30th June 2015, the mean pressure is 988. The coefficient of pressure correct (β) is -0.9295[12]. The pressure-corrected muon hourly counts are shown at Fig.2. Because the complexity of atmospheric environment, the different of pressure-corrected counts with no corrected counts is not evident.

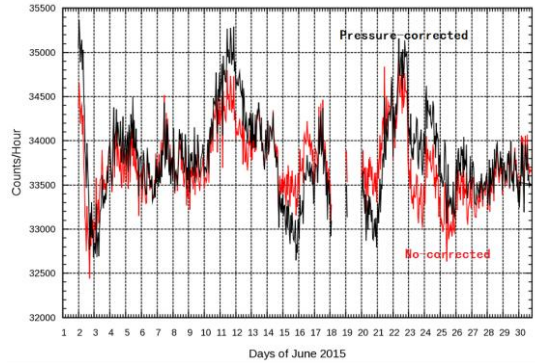

Fig. 2　Pressure-corrected muon hourly counts observed by theZSMT from 1st to 30th June 2015.

## 3　Observations for Forbush decrease in June 2015

An impressive series of solar events occurred in June 2015. The Earth experienced a geomagnetic storm on June 22, 2015 due to the arrival of an Earth-directed CME from June 21. The CME originated at 2:36 UT on June 21[13]. Coronal material exploded from the sun with the velocity of 1366 km/s, arriving at the Earth at 5:45 UT on June 22. The same active region produced two other CMEs in the past few days, which were pushed (or compressed) along by the faster Earth-directed CME from June 21. The sudden storm commencement (SSC) identifies itself with the arrival of the ICME shock at 18:33 UT on June 22. This ICME are recorded from 02:00 UT at 23 on June to 14:00 UT on 24 [14]. A shock arrived at 13:29 UT on June 24 with an ICME from 25 June at 10:00 UT to 26 June at 06:00 UT. This ICME is related to the halo CME that was released at 18:36 UT on June 22.

During these fast ICMEs directed earthwards, a large FD event happens. We examine the sequence of solar events that occurred during 21-25 June and how they affect the counting rates of the ZSMT.

### 3.1　The galactic cosmic ray data

We study the characteristics of the FD using neutron monitor (NM) data and muon telescope (MT) data for 22-26 June 2015.

The pressure corrected hourly rates (%) from selected neutron monitors of the Bartol Research Institute are plotted in Fig.3. The South Pole neutron monitor is located locates at 90S. The McMurdo neutron monitor is located locate at 77.9S 166.6E. The mean value take from 1st to 20th, before IMF distribution. It is evident that the hourly rates for NMs in the Antarctic region, include the South Pole and McMurdo, shows evident FD onset (18:00 UT) before the shock arrival (SSC) at 18:33 UT on June 22 with 0.5 hours in advance. Each of the NMs shows very similar variations. Two neutron monitors began to decrease with the arrival of the shock and reached the minimum at 11:00 UT on June 23 while the Earth entered in the ICME. In association with a shock arrived at 13:29 UT on 24 and an ICME, this ICME is related to the CME that was released at 18:36 UT on June 22, the two neutron monitors decreased the counting rate again. On the other hand, a slight difference was recognized during decrease between the McMurdo and South Pole. The McMurdo had a pre-increase before the shock arrival while the South Pole had not. During the decrease time, the McMurdo shows an evident four-step decrease while the South Pole data did not show any obvious effect.

Fig.3 shows a plot of the hourly rates (%) in the Vertical (V) for Nagoya muon telescope (MT). The hourly counting rate of the Nagoya muon telescope was taken from the Global Muon Detector Network (GMDN) data system of Shinshu University. The mean value was evaluated take from 1st to 20th day before, of the IMF distribution. The Nagoya MT locates at 35.2N 137E. The $R_m$ in the Vertical (V) values is 67 GV. The plot for Nagoya MT indicates that FD onset (19:00 UT) delayed (～0.5h) t to the arrival of the SSC (18:33 UT) and reached the minimum at 7:00 UT while the Earth entered in ICME. A pre-increase is visible obviously before FD onset. In association with a shock arrival at 13:29 UT on 24 and an ICME, the hourly counts of the Nagoya MT did not decrease again in comparison with the data of the NMs of South Pole and McMurdo.

As described in Section 2.3, the pressure change is very strong at Zhongshan station. It is necessary to a pressure correct of muon data for ZSMT. Because the count rate varies sharply as shown in Fig.5, an hourly count 34300, an hourly count before IMF distribution on June 22, is taken as the hourly mean value. The cosmic ray hourly rates (%) after pressure corrected of ZSMT is displayed in Fig.3. The plot for ZSMT indicates that FD onset time (21:00 UT) is delayed (～2.5h) with respect to the SSC (18:33 UT) and reach the minimum at 7:00 UT while the Earth entered the ICME. The counts had a pre-increase in the leading to FD onset; in association with a shock arrivaled at 13:29 UT on 24 and an ICME, the hourly counts of the ZSMT decreased simultaneous with the counting rate of the NMs of the South Pole and McMurdo.

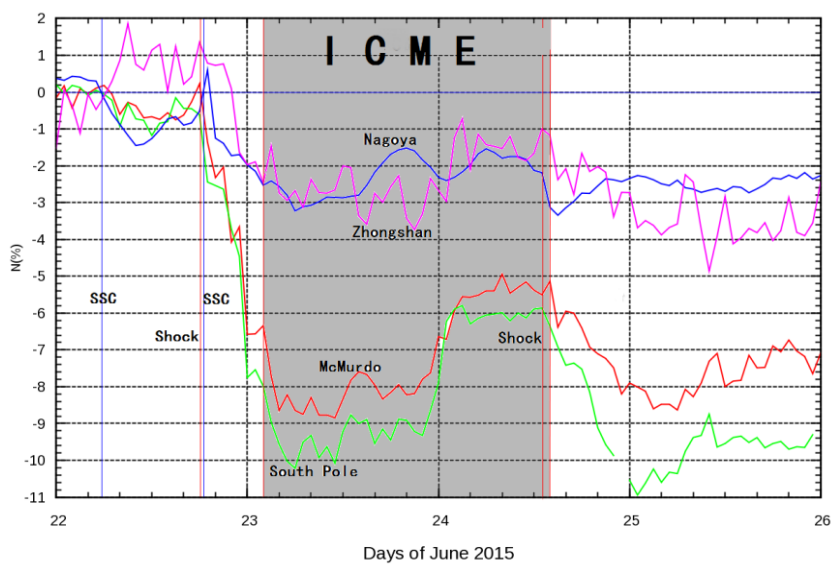

Fig. 3 The cosmic ray intensity variation (%) hourly values of neutron monitor data and muon telescope. Data are plotted for June 22-26, 2015.

**3.2 The solar wind measurements**

From June 21 to 25, 2015, a large FD event happens. The solar wind measurements from ACE and WIND are shown in Fig.4. The figure show the magnetic field intensity B, magnetic field $B_x$, $B_y$, $B_z$ (in the GSE coordinate system) component, solar wind proton density $n_p$, solar wind proton temperature $T_p$, the solar wind proton speed $V_p$, and the geomagnetic activity index Dst.

The interplanetary shocks observed by the wind spacecraft: CfA Interplanetary Shock Database - Yearly Summary [15]. The times of shock is shown by solid vertical line. The time of the ICME associated geomagnetic storm sudden commencement (SSC), typically related to the arrival of a shock at Earth, take from the Service International des Indices Geomagnetique[16] . The time of SSC is shown by dashed vertical line.

The ICME contents from the list: Near-Earth Interplanetary Coronal Mass Ejections Since January 1996 compiled by Ian Richardson and Hilary Cane [14]. The ICME start time at 20200 UT of 23 and end at 1400 UT of 24. The gray shaded area indicates the interval of an ICME. The ICME shows an evidence of a rotation of the field direction, but lacks some other characteristics of a magnetic cloud, for example an enhanced magnetic field. Mean speed of the ICME speed is 610 km/s, based on solar wind speed observations during start and end times of the ICME. Maximum solar wind speed is 740 km/s during the period from the disturbance of the shock to the trailing edge of the ICME. If the magnetic field within the ejecta is organized as a flux rope, then the magnetic field can be fitted by Higalgo's model [17]. By only having a quick look at the graph (Fig.4), we have found that this ejecta without any kinds of well-defined topology (no clear signatures: neither in the magnetic field components, which present a complex profile, nor in the plasma behavior). The profiles of the magnetic field component do not present the smooth and regular behavior expected for a flux-rope topology, but showed large fluctuating profiles during the interval.

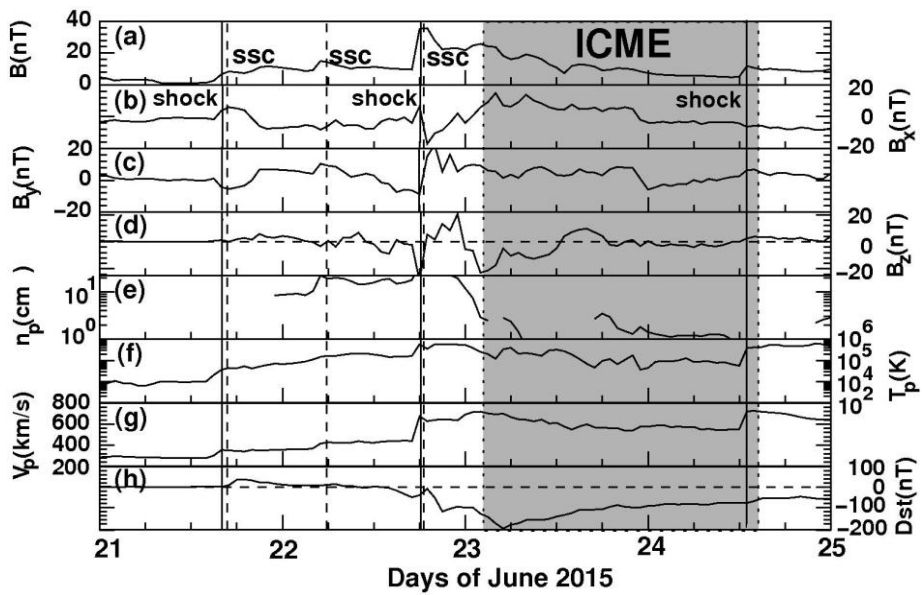

Fig.4 Data are plotted for June 21-25, 2015.The figure from top to bottom shows hourly values of (a) the magnetic field strength B(nT) in the GSE coordinate system, (b) the magnetic field component $B_x$, (c) the magnetic field component $B_y$, (d) the magnetic field north-south component $B_z$, (e) the solar wind proton density $n_p$, (f) the solar wind proton temperature $T_p$, (g) the solar wind proton speed $V_p$, (h) the geomagnetic activity index Dst. The start times of shock and SSC are shown by solid and dashed vertical lines, respectively. The shade gray area indicates the interval of an ICME.

To study the fine structure of salient features, Fig.5 shows hourly plots of the magnetic field intensity $B_Z$, the solar wind proton speed $V_P$, the geomagnetic activity index Dst, the geomagnetic $K_P$ index and the intensity variation profiles of the ZSMT cosmic rays N(%) with an expanded time scale for 24 hours (12UT June 22 – 24 UT June 24, 2015). The positions of the shock and sharp storm sudden commencement (SSC) are marked; ICME lies to the right of the SSC (the shade gray area). The time of SSC (typically related to the arrival of a shock at Earth) is 22 18:33 UT.

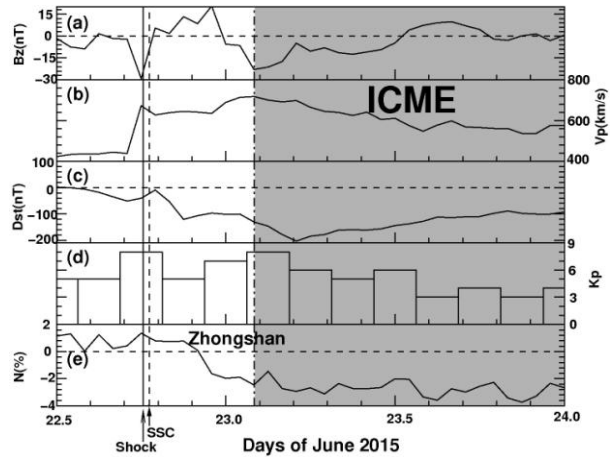

Fig.5 Data are re-plotted for 12 UT June – 18 UT June 24, 2015. The figure from top to bottom shows hourly values of (a) the magnetic field strength $B_Z$(nT) in the GSE coordinate system, (b) the solar wind proton speed $V_p$, (c) the geomagnetic activity index Dst, (d) $K_p$ index, (e) the cosmic ray intensity variation(%) profiles of ZSMT The start times of shock and SSC are shown by solid and dashed vertical lines, respectively. The shade gray area indicates the interval of an ICME

The important interplanetary parameter for 'GCR-effectiveness' appears to be the enhanced and turbulent magnetic field. Scattering of cosmic-ray particles by an enhanced turbulent magnetic field in the sheath region between the shock front and the CME or magnetic cloud appears to be the most effective mechanism to produce FD in cosmic rays [18]. A sharp rise in the counts (N) followed by a steep increase in $B_Z$ (Fig.5) and $V_P$ when the Earth enters the region between the SSC with leading edge of the ICME leading to the FD onset; this may be the region of turbulence. A sharp recovery occurs as the earth entrance in the ICME, a low GCR density region.

**3.3 Enlil simulations of the June CME**

We now focus in particular on the 2015 June 21 CME. Using the WSA-ENLIL model, which is a large-scale physics-based prediction model of the heliosphere, we proceed with investigating the heliospheric effect of the CMEs. Ambient solar wind conditions were initiated from photospheric magnetograph data using the Wang-Sheeley-Arge model [19]. To simulate the passage of the CME, a "cone" model of the CME was used to provide the kinematic properties of a transient density plug that was introduced into the magnetogram-derived ambient solar wind simulation [20]. Figure 6 shows two snapshots of the Enlil model run at 18 UT on 22 when the June 21 CME leading edge was at 1 AU. The model results show density and velocity of the solar wind in the ecliptic plane, where blue color represent low density and red/white/gray colors represent high densities (velocity) of the solar wind [21]. The CME shock arrival time is: 2015-06-22T17:59Z. The simulation predicted CME shock arrival time is: 2015-06-22T21:43Z. Enlil simulations of the June 18 and 19 CMEs were also performed, though are not shown here.

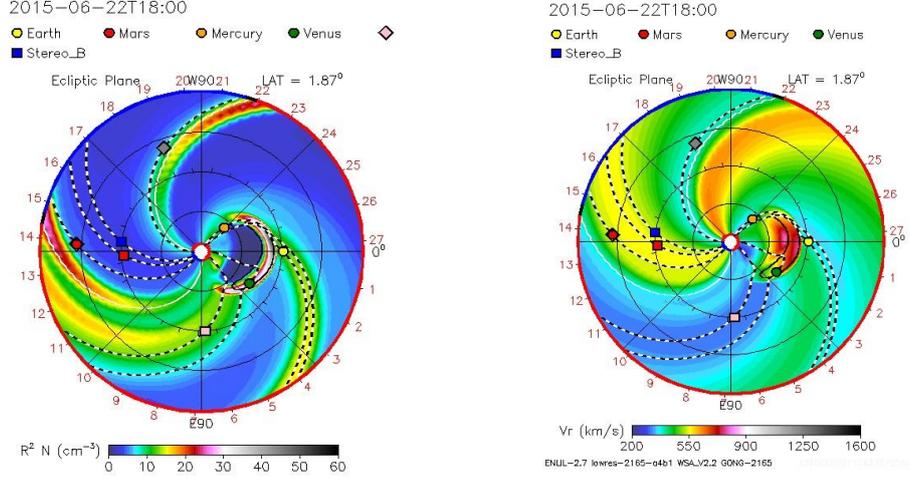

Fig. 6 Snapshots of the Enlili model run at 18 UT on 22 during the 2015 June 21 CME's propagation through the inner heliosphere [21]. The figures shown density (left) and velocity (right) of the solar wind. The Sun is marked as the white circle and the Earth is the yellow circle. Time is expressed in universal time (UT).

## 4  Summary

At Zhongshan Station in Antarctic, a cosmic ray muon telescope was installed in 2004. An impressive series of solar events occurred in June 2015. From June 21 to 25, 2015, a large FD event happens. We study the characteristics of the FD using muon telescope (MT) data and neutron monitor (NM) data. The counting rates of the ZSMT indicate that FD onset (21:00 UT) delayed (～2.5h) with respect to the SSC (18:33 UT) and reached the minimum at 7:00 UT when the Earth entered in the ICME. In association with a shock arrival at 13:29 UT on 24 June and an ICME, the hourly counting rate of the ZSMT decreased simultaneously with NMs of South Pole and McMurdo. The FD onset (19:00 UT) of Nagoya MT delayed (～0.5h) with respect to the SSC (18:33 UT) . These results is different from the FD onset time of Nagoya MT that delayed (～5h) with respect to the SSC during 14 and 18 May, 2005[22]. A pre-increase before the shock arrival also was observed. The traditional model of FDs predicts that an ICME and its shock make reduction of the GCR intensity with a two-step profile [23].   However, this FD had a profile of four-step decrease. The traditional one- or two-step classification of FDs was inadequate to explain our study.   Each FD must be studied as a unique event in the detailed context of its driving interplanetary conditions [24]. It was a four steps FD, probably due to the interaction of the faster CME on 21 June and the two slow CMEs in the past few days, which were pushed along by the faster the Earth-directed CME from June 21. The first step was caused by the shock. The second and the third one were caused by the entry into the extended sheath region. Finally, CRs entered in the enhanced region of the ejecta [25]

Due to a very small detection area, ZSMT counting rate fluctuations are too large. It is further work to extend detection area. By the end of 2015, the ZSMT was enlarged to 1 m ✕ 1m that consists of 4 four 0.5 m × 0.5 m detectors. In addition, the observation data has achieved by satellite link real-time transmission, allowing real-time data exchange between Zhongshan station and cosmic ray observatories worldwide (for example, NMDB (real-time database for high resolution neutron monitor measurements) and SPACESHIP EARTH), which promotes the joint study of the space environment.

*Acknowledgements:*

*We would like to express our appreciation to the staff ( Liu Yang and Shen Xin) of the Zhongshan station for their valuable work. We thank the ACE SWEPAM instrument team and the ACE Science Center for providing the ACE data. The neutron-monitor data are provided by the University of Delaware department of Physics and Astronomy and the Bartol Research Institute with supported from the U.S. National Science Foundation under grant ANT-0838839. We thank the Cosmic Ray Experimental Science Team for using the Global Muon Detector Network (GMDN) data of Shinshu University. The results presented in this paper rely on geomagnetic indices calculated and made available by ISGI Collaborating Institutes from data collected at magnetic observatories. We thank the involved national institutes, the INTERMAGNET network and ISGI (isgi.unistra.fr). We thank the Community Co-ordinated Modeling Center for the ENLIL model run result.*